\documentclass[english]{report}

\usepackage{graphicx}
\usepackage{astro_bib_macro}
\usepackage{txfonts}
\usepackage{natbib}        
\bibpunct{(}{)}{;}{a}{}{,} 

\begin{document}

\begin{center}
{\Large A close-in companion to HD142384\footnote{Based on observations collected with the PIONIER/VLTI instrument at the European Southern Observatory, Paranal, Chile.}}
\\ \vspace{0.5cm} 
J.-B.~Le~Bouquin \\
  \textit{Univ. Grenoble Alpes, IPAG, F-38000 Grenoble, France}  \\
  \textit{CNRS, IPAG, F-38000 Grenoble, France}
\\ \vspace{0.5cm} 
Prepared 2014/06/25
\end{center}

\noindent {\bf Abstract:} I report the detection of a companion around HD142384 from Optical Long Baseline Interferometry in the near-infrared (flux difference $\Delta H\mathrm{mag}\approx4$ and separation $\approx20\,$mas). HD142384 has been used as PSF reference for published observations of the well-known young star HD142527. This is now strongly discouraged.

~\newline

\section*{Observations}
\label{sec:observation}

Over the last year, HD142384 has been used as reference star for various observations performed with the PIONIER combiner \citep[][]{Le-Bouquin:2011}  from the  Very Large Telescope Interferometer \citep[VLTI,][]{Haguenauer:2010}. By chance, most of these observations were double-calibrated by at least one additional reference star.

Table~\ref{tab:obs} summarizes the date of observations and the list of additional reference stars that could be used to calibrate the observations of HD142384. These reference stars were found with the \texttt{SearchCal}\footnote{http://www.jmmc.fr/searchcal\_page} \citep{Bonneau:2011a}. \emph{A-priori}, these reference star are not more trustable than HD142384 and we are blocked in a chicken-and-egg problem. However two arguments support the idea that HD142384 is the culprit:
(1) During good nights, the other reference stars provide consistent transfer function estimates.
(2) The transfer function measured on HD142384 is systematically \emph{lower} than those computed with the other reference stars.

\begin{table}[h]
\caption{Log of observations with the reference stars.}
\centering
\begin{tabular*}{0.95\columnwidth}{ll}
\hline\hline\noalign{\smallskip}
Date  & Reference stars \\
\noalign{\smallskip}\hline\noalign{\smallskip}
2013-02-19 & HD142132 \\
2013-04-21 & HD144988 \\
2013-06-03 & HD143561, HD144073, HD144475 \\
2013-06-06 & HD142132, HD175714\\
2013-06-14 & HD144475, HD142132\\
2013-06-15 & HD142132 \\
2013-06-16 & HD142132, HD137598 \\
2013-07-02 & HD142132 \\
2014-05-28 & HD142132, HD144475 \\
2014-06-02 & HD142132, HD144475 \\
2014-06-21 & HD142132, HD144475, HD146906 \\
\noalign{\smallskip}\hline
\end{tabular*}
\label{tab:obs}
\end{table}

\section*{Binary fitting}

The calibrated observations of HD142384 can be satisfactory adjusted with a binary model. We assume a diameter of $0.5\,$mas for the primary, according to \texttt{SearchCal}, and consider that the secondary is point-like. The best fitting flux ratio is around 2.5\%, that is $\Delta H\mathrm{mag}\approx4$. The best fit binary models for each epoch are shown at the end of the manuscript.

Given the separation and flux ratio, the probability of a contamination by a background /foreground object is extremely small. The orbital motion is detected over one year. This is well compatible with the physical size of the separation (nearly 2AU) according to the Hipparcos parallaxe $\pi=7.49$\,mas.

\section*{Discussion}

HD142384 has been used as reference by \citet{Biller:2012} which reports the detection of a low-mass stellar companion to the transitional disk star HD142527 using Spare Aperture Masking (SAM/VLT). HD142384 is the only reference that was observed with the H-band filter. The new companion detected around this reference star (HD142384B, 2.5\% in H-band) is significantly brighter than the tentative detection of HD142527b (1.6\% in H-band, 0.8\% in L-band).

Given its close-in separation, HD142384B has probably insignificant influence on the calibration performed in L-band where the spatial resolution of SAM/VLT is $90$\,mas. This is confirmed by the trend shown in Fig.~1 from \citet{Biller:2012}, where all the three reference stars are consistent. In H-band, the situation is not as clear. The spatial resolution of SAM/VLT is $40\,$mas, that is only twice the possible separation of HD142384B.

Alltogether, the use of HD142384 as reference star is now strongly discouraged, especially with SPHERE or GPI.

\subsection*{Acknowledgments}
{\it \small PIONIER is funded by the Universit\'e Joseph Fourier (UJF), the Institut de Plan\'etologie et d'Astrophysique de Grenoble (IPAG), the Agence Nationale pour la Recherche (ANR-06-BLAN-0421 and ANR-10-BLAN-0505), and the Institut National des Science de l'Univers (INSU PNP and PNPS). The integrated optics beam combiner is the result of a collaboration between IPAG and CEA-LETI based on CNES R\&T funding. This works made use of the Smithsonian/NASA Astrophysics Data System (ADS) and of the Centre de Donnees astronomiques de Strasbourg (CDS). All calculations and graphics were performed with the freeware \texttt{Yorick}.}

%
\renewcommand{\chapter}[2]{}

\begin{figure}[!h]
  \centering
\includegraphics[width=0.8\textwidth]{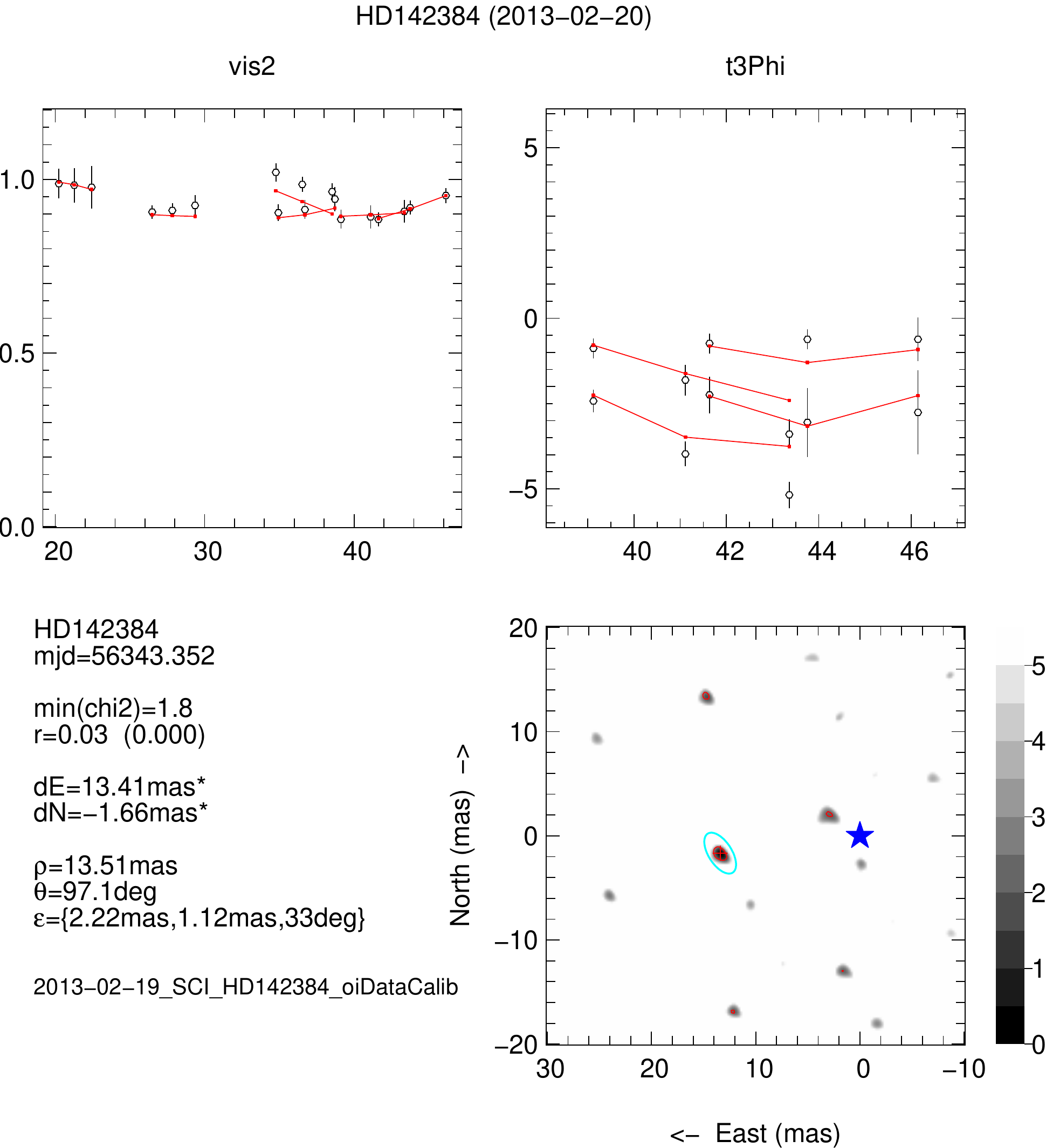}
\label{fig:fit}
\end{figure}
\begin{figure}[!h]
  \centering
\includegraphics[width=0.8\textwidth]{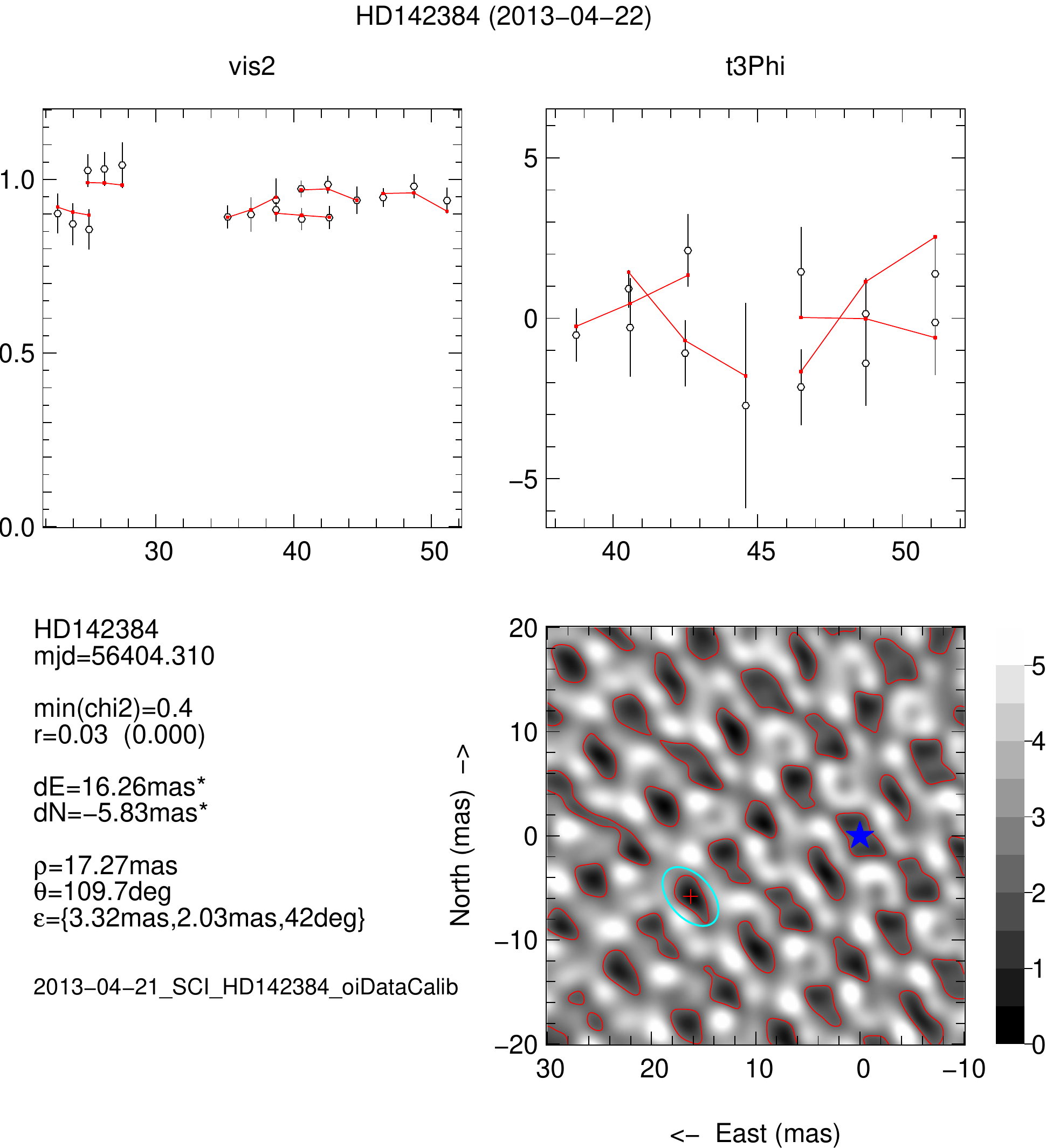}
\includegraphics[width=0.8\textwidth]{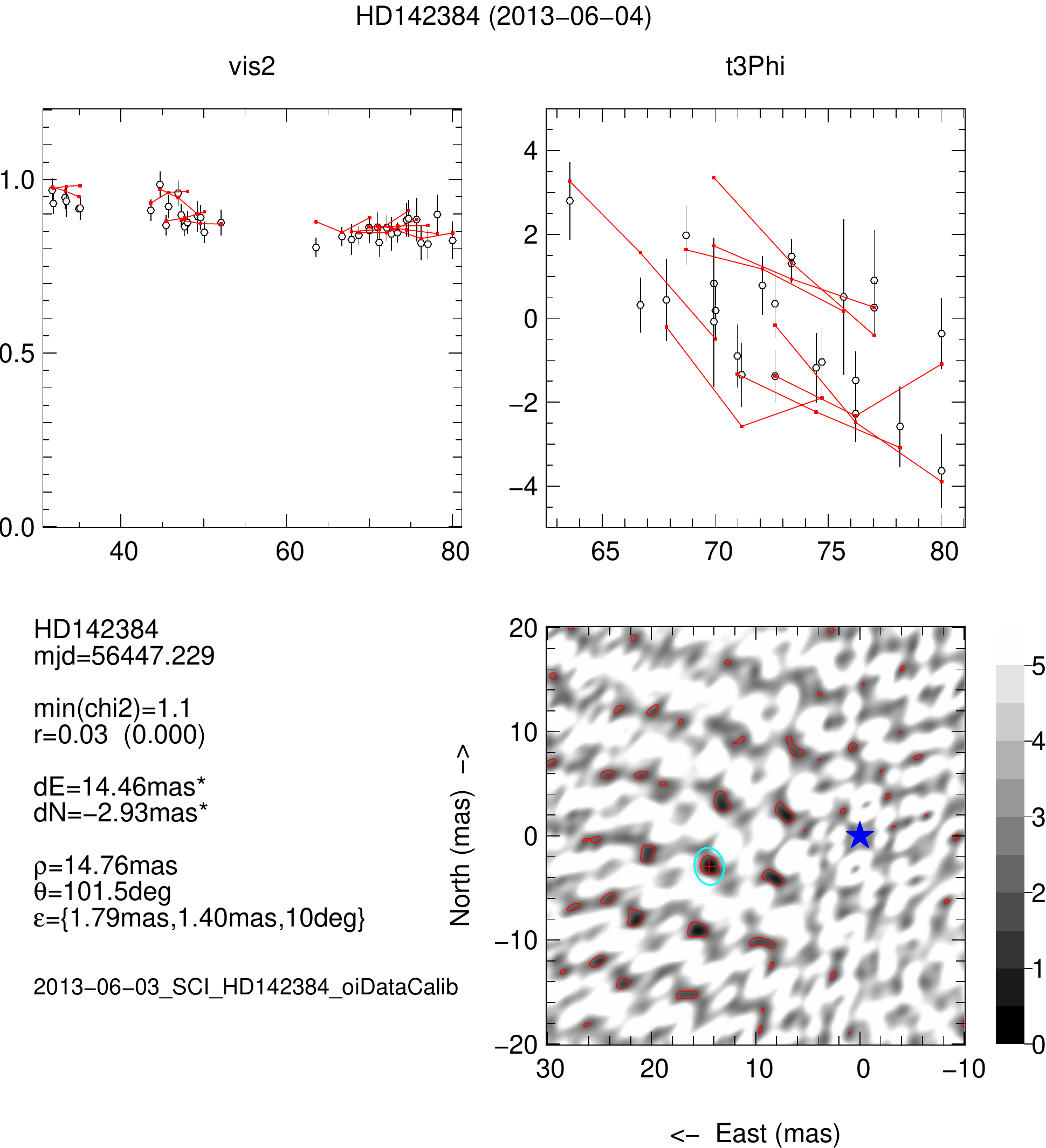}
\label{fig:fit}
\end{figure}
\begin{figure}   \centering
\includegraphics[width=0.8\textwidth]{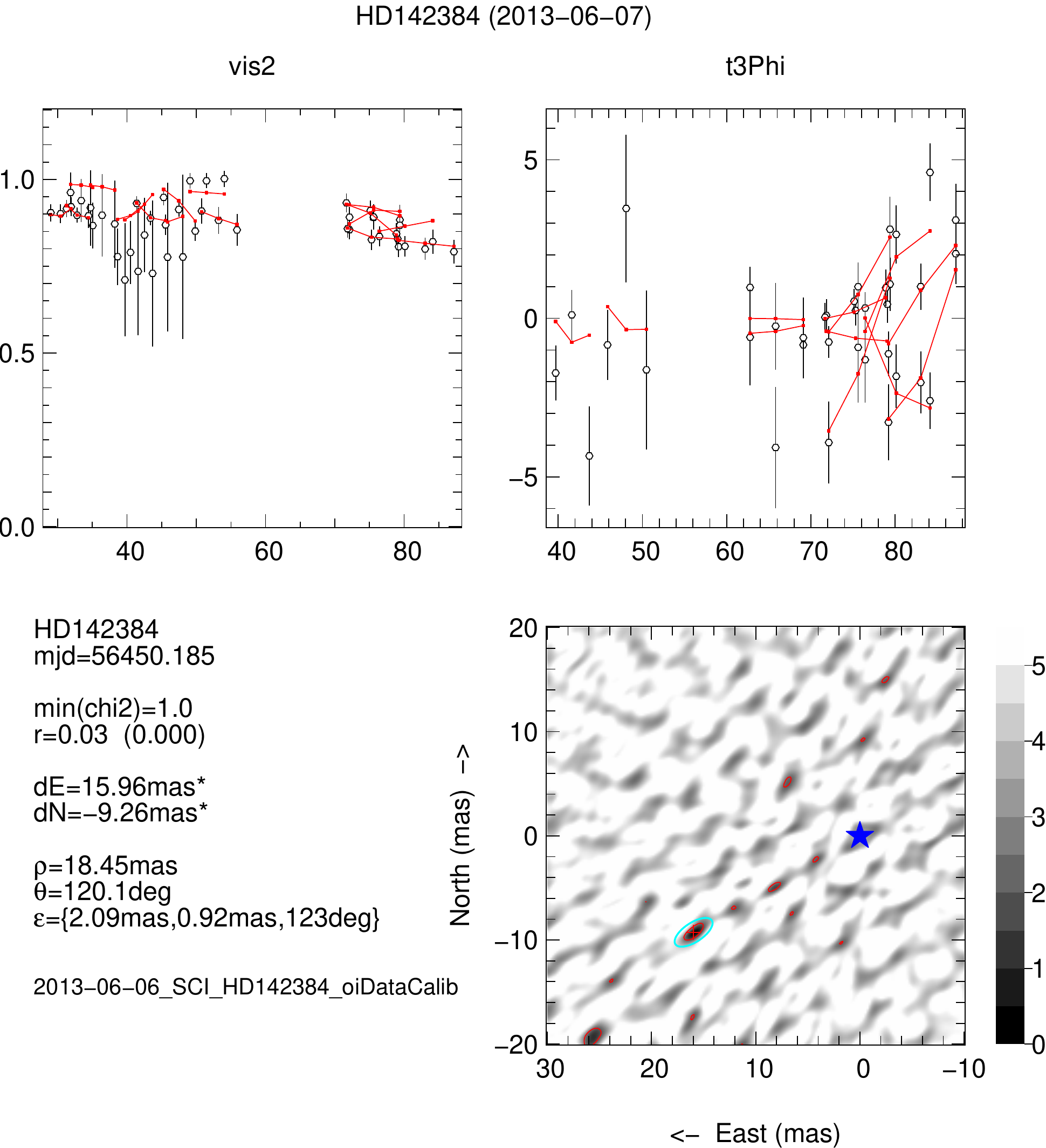}
\includegraphics[width=0.8\textwidth]{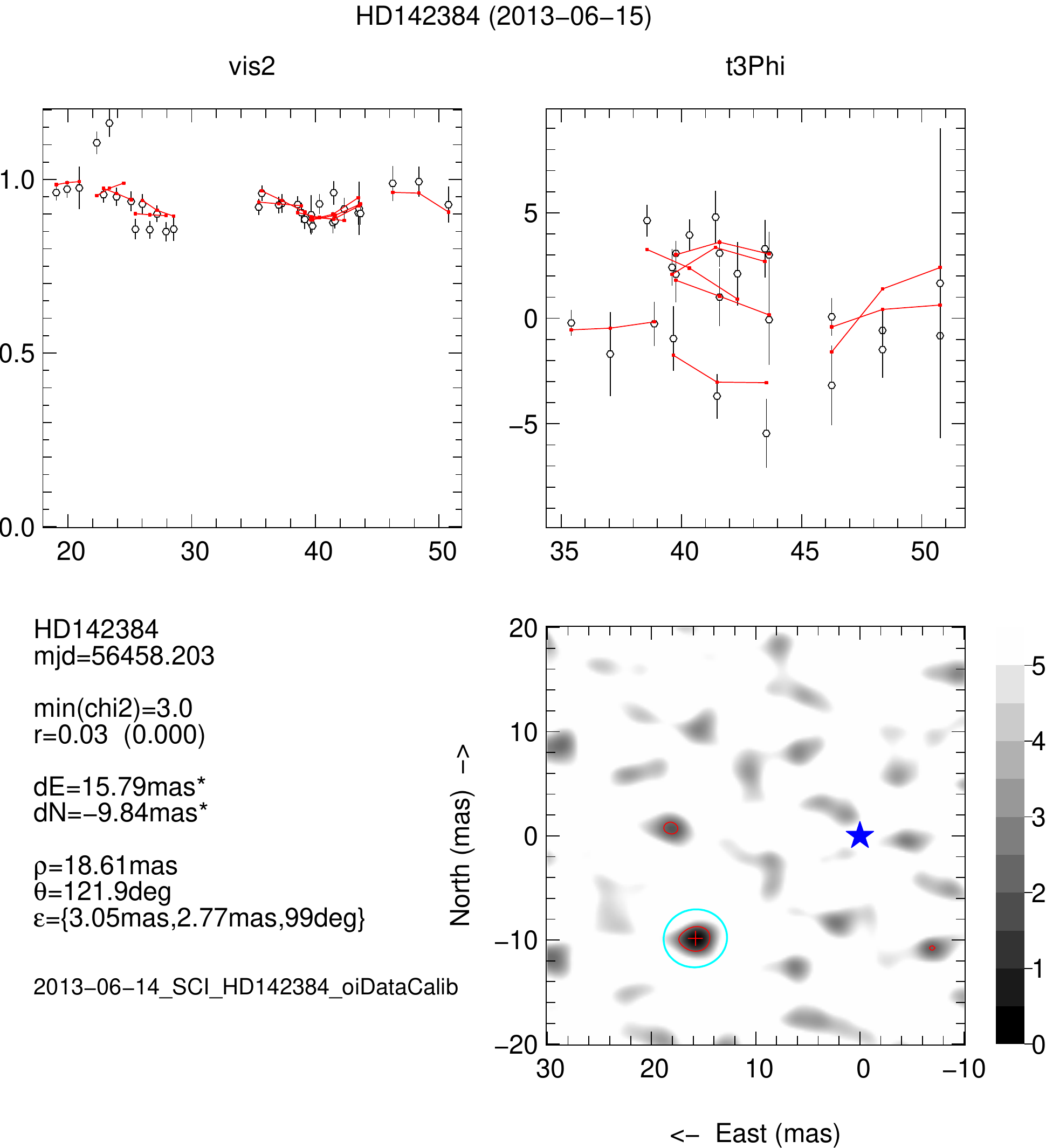}
\end{figure}
\begin{figure}   \centering
\includegraphics[width=0.8\textwidth]{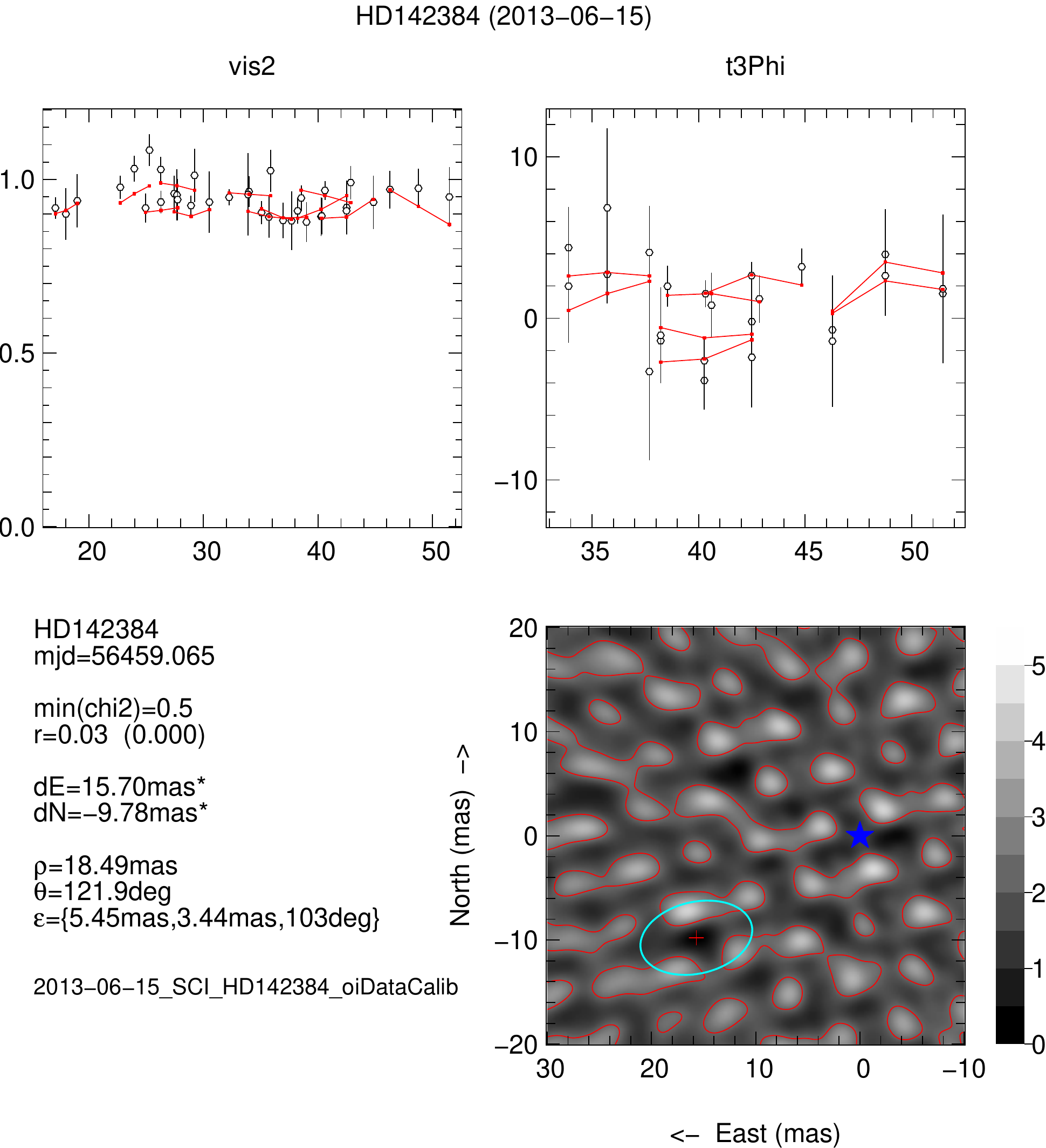}
\includegraphics[width=0.8\textwidth]{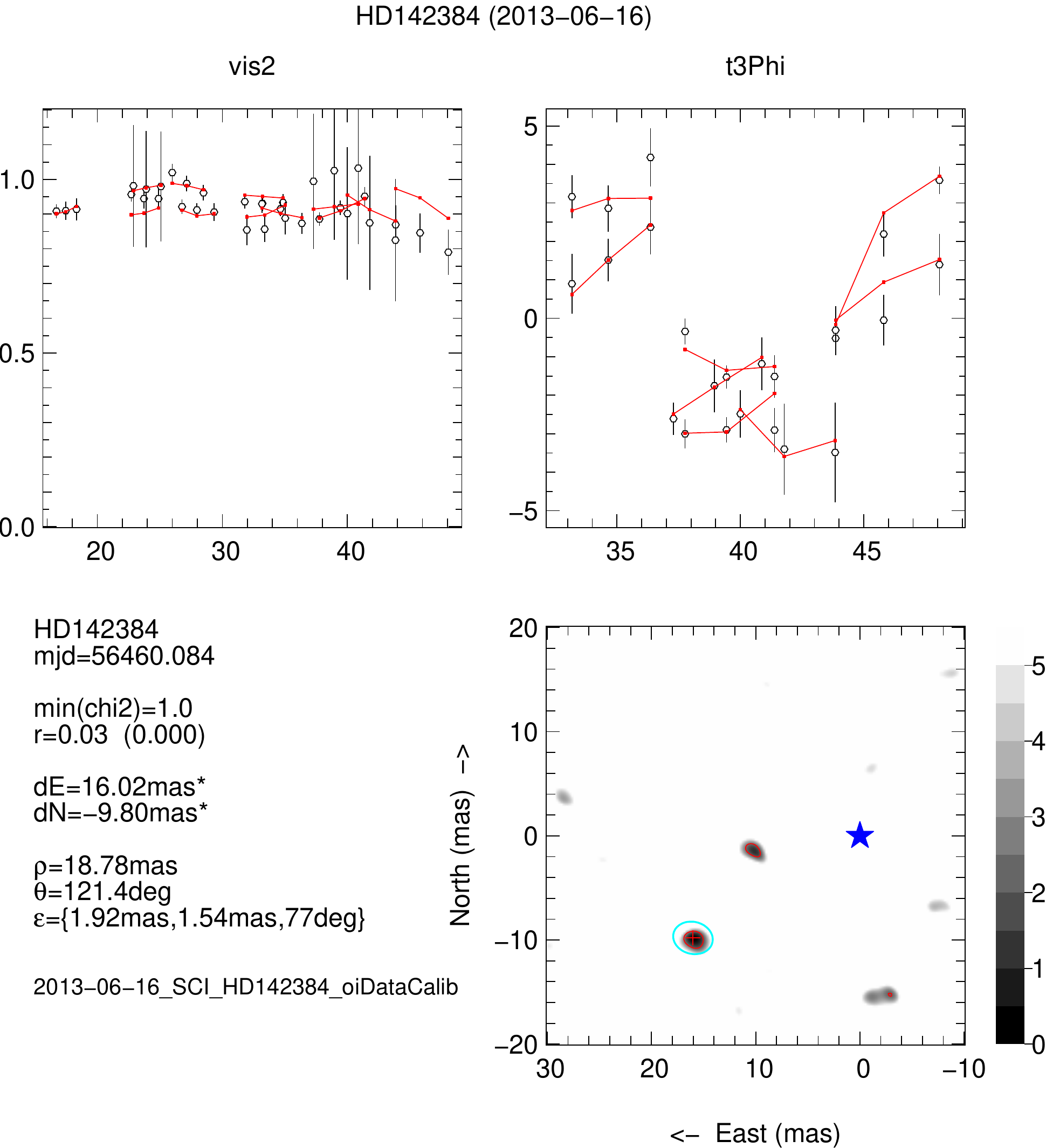}
\end{figure}
\begin{figure}   \centering
\includegraphics[width=0.8\textwidth]{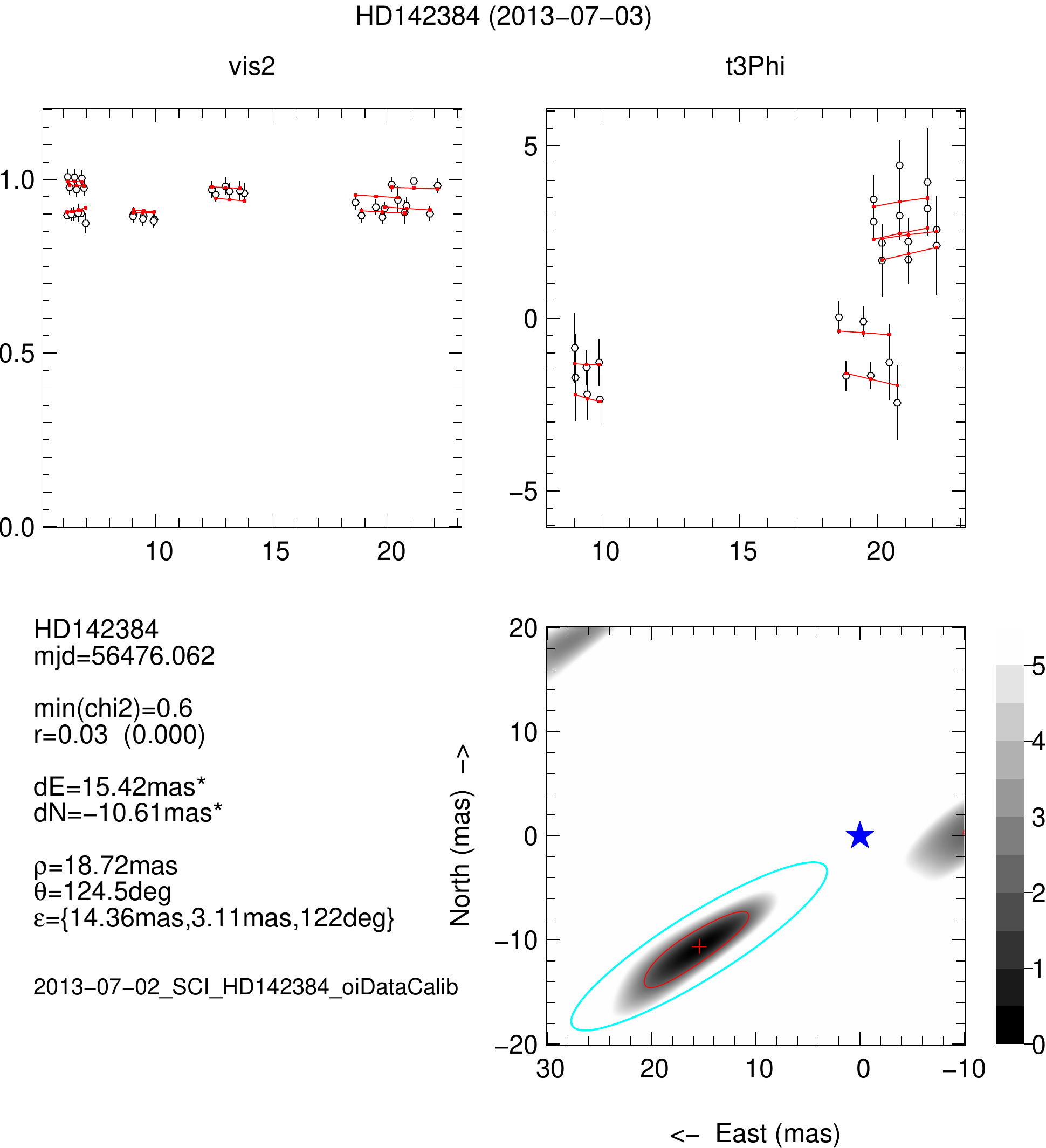}
\includegraphics[width=0.8\textwidth]{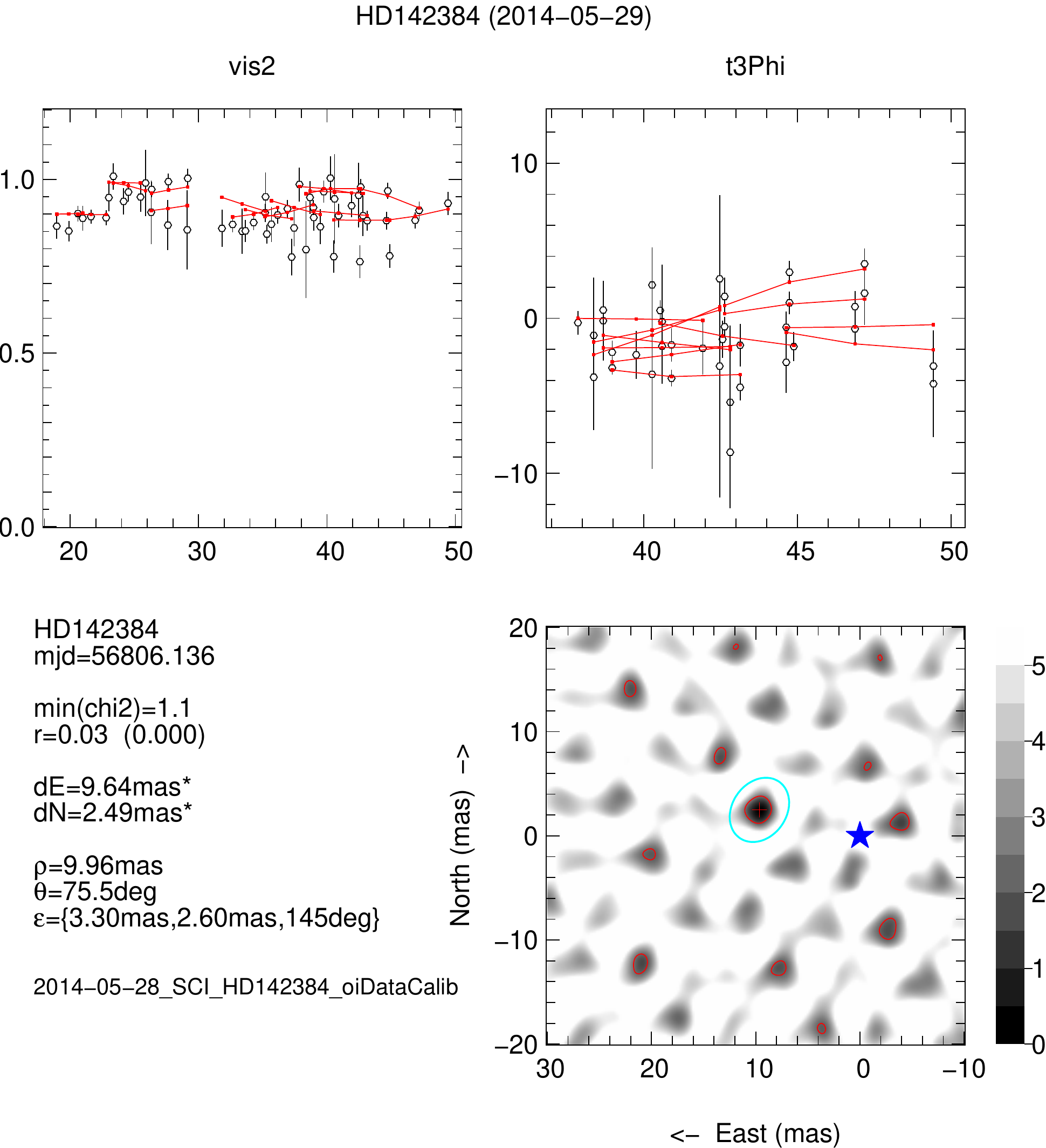}
\end{figure}
\begin{figure}   \centering
\includegraphics[width=0.8\textwidth]{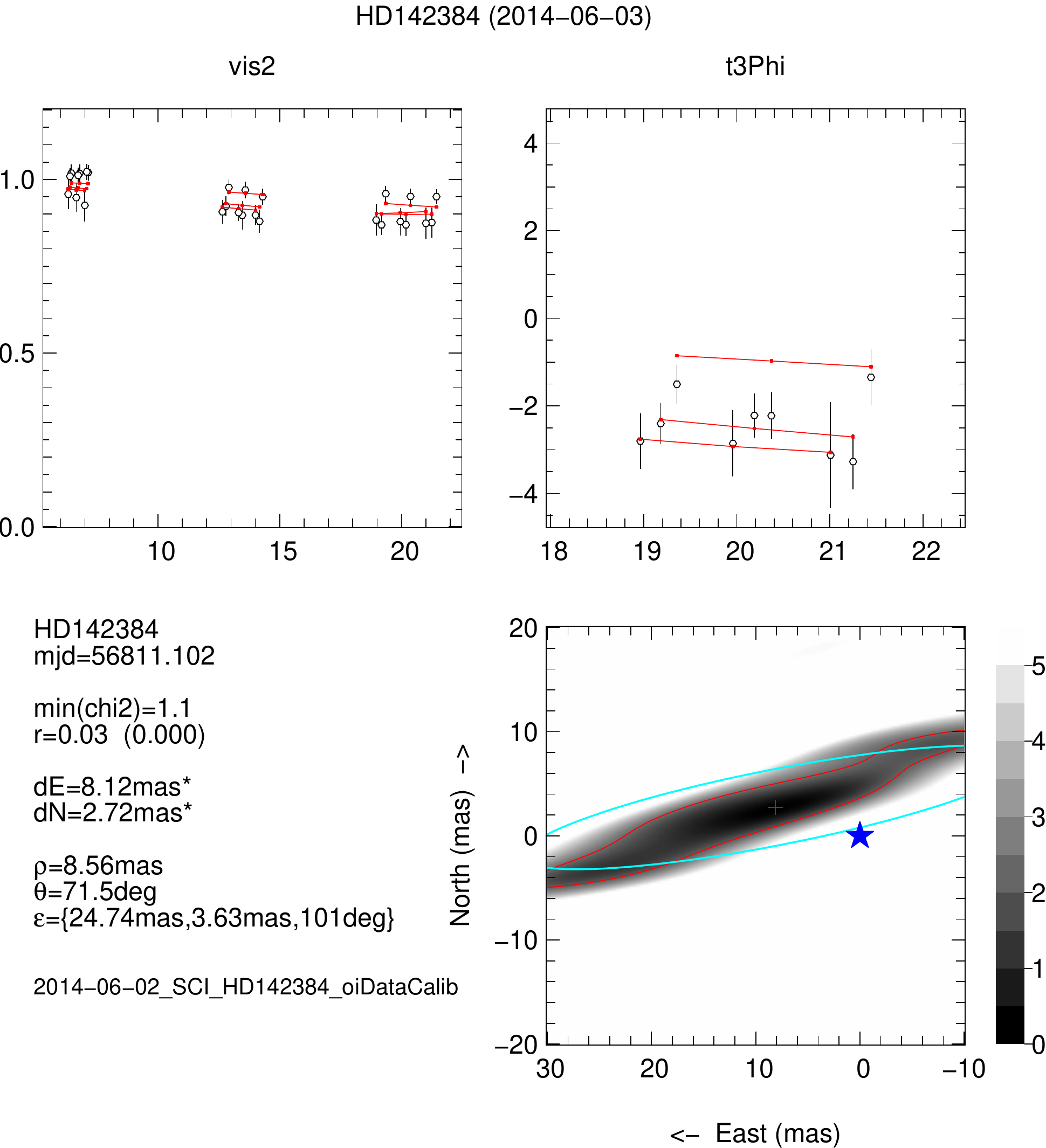}
\includegraphics[width=0.8\textwidth]{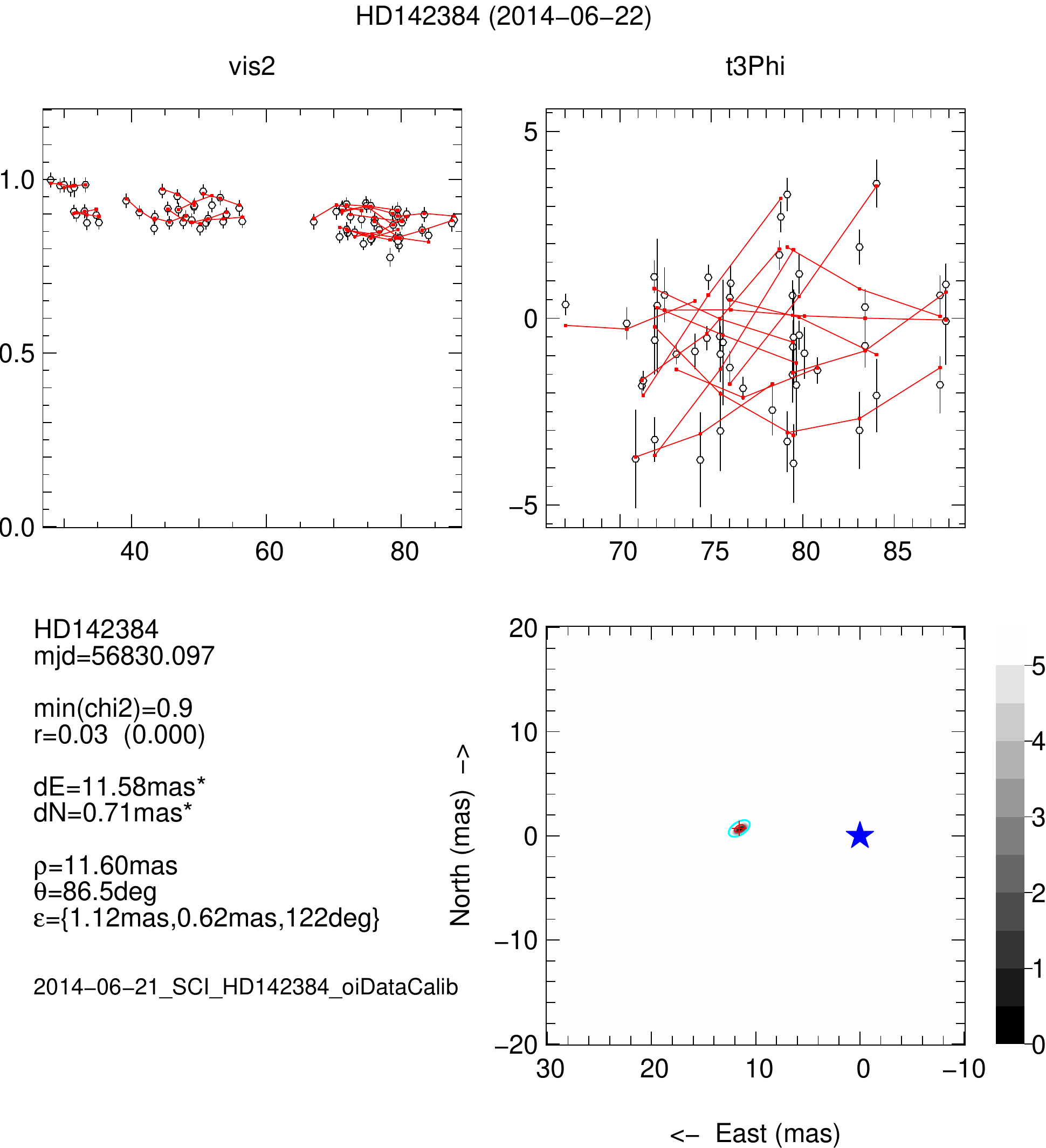}
\end{figure}

\end{document}